\documentclass[a4paper]{article} 
\usepackage{spconf,amsmath,epsfig,graphics,url}

\usepackage{amssymb}
\usepackage[font=footnotesize, belowskip=-6pt]{caption}
\usepackage{graphicx,subcaption} 
\usepackage[hidelinks]{hyperref}
\usepackage[hang,flushmargin]{footmisc}

\newcommand\blfootnote[1]{%
  \begingroup
  \renewcommand\thefootnote{}\footnote{#1}%
  \addtocounter{footnote}{-1}%
  \endgroup
}

\usepackage{fancyhdr}
\usepackage{etoolbox}
\patchcmd{\thebibliography}
  {\settowidth}
  {\setlength{\parsep}{0pt}\setlength{\itemsep}{0pt plus 0.1pt}\settowidth}
  {}{}

\title{Interactive Light Field Tilt-Shift Refocus with Generalized Shift-and-Sum}
\name{Martin Alain\textsuperscript{*}, Weston Aenchbacher\textsuperscript{*}, Aljosa Smolic \vspace{-0.2cm}}
\address{V-SENSE Project, School of Computer Science and Statistics, Trinity College, Dublin}

\begin{document}
\ninept
\urlstyle{same}


\twocolumn[{
\renewcommand\twocolumn[1][]{#1}
\maketitle
\begin{center}
\centering
\vspace{-0.7cm}
    \includegraphics[width=.32\textwidth]{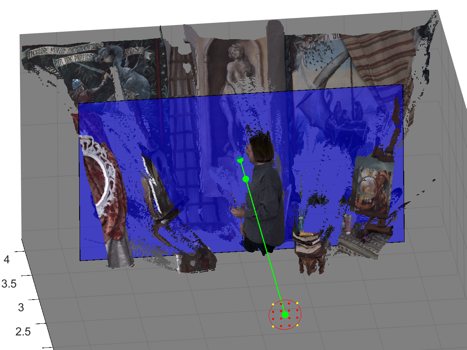}
    \hfill
    \includegraphics[width=.32\textwidth]{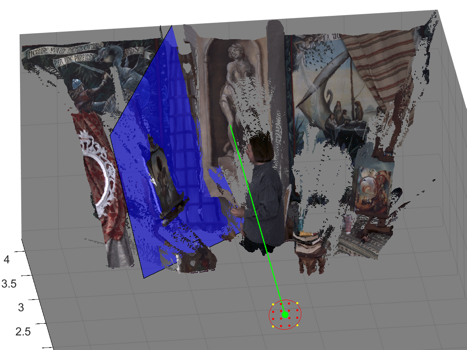}
    \hfill
    \includegraphics[width=.32\textwidth]{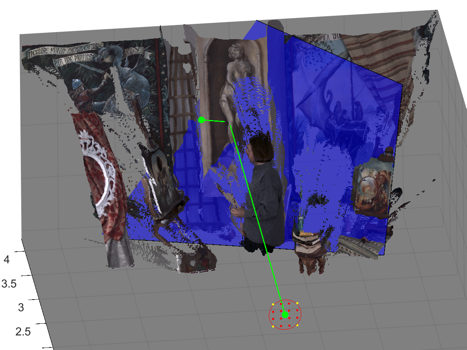}
    
    \vspace{0.5mm}
    \includegraphics[width=.32\textwidth]{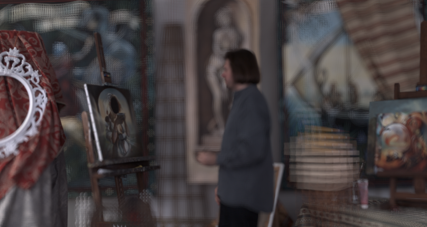}
    \hfill
    \includegraphics[width=.32\textwidth]{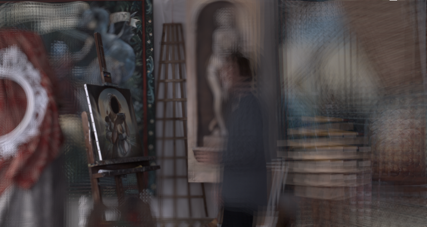}
    \hfill
    \includegraphics[width=.32\textwidth]{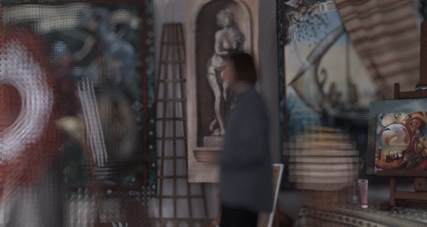}
    \setlength{\belowcaptionskip}{0pt}
    \captionof{figure}{Interactive tilt-shift refocus results on a frame of the ``Painter'' light field video, from the Technicolor dataset [3]. \textbf{(top)} Point clouds of the scene geometry with the refocus plane in transparent blue, camera positions in red and yellow, and optical axis and normal vector in green. \textbf{(bottom)} Tilt-shift refocus images for each of the refocus plane positions shown above. \textbf{(left)} Frontoparallel refocus. This is the same result that could be achieved with existing shift-and-sum but is implemented in the tilt-shift framework. \textbf{(center)} Refocus plane aligned with the painting on the left. \textbf{(right)} Refocus plane aligned with the painting on the right. Image size has been reduced in this pre-print.}
    \label{fig:Painter}
\end{center}
}]

\begin{abstract}
Since their introduction more than two decades ago, light fields have gained considerable interest in graphics and vision communities due to their ability to provide the user with interactive visual content. One of the earliest and most common light field operations is digital refocus, enabling the user to choose the focus and depth-of-field for the image after capture. A common interactive method for such an operation utilizes disparity estimations, readily available from the light field, to allow the user to point-and-click on the image to chose the location of the refocus plane.

In this paper, we address the interactivity of a lesser-known light field operation: refocus to a non-frontoparallel plane, simulating the result of traditional tilt-shift photography. For this purpose we introduce a generalized shift-and-sum framework. Further, we show that the inclusion of depth information allows for intuitive interactive methods for placement of the refocus plane. In addition to refocusing, light fields also enable the user to interact with the viewpoint, which can be easily included in the proposed generalized shift-and-sum framework.\blfootnote{\textsuperscript{*}Authors contributed equally to the publication.

This project has received funding from the European Union’s Horizon 2020 research and innovation
programme under grant agreement No 780470.

This publication has emanated from research conducted with the financial support of Science Foundation Ireland (SFI) under the Grant Number 15/RP/2776.}
\vspace{-0.4cm}
\end{abstract}


\section{Introduction}
\label{sec:Intro}
\vspace{-0.21cm}

Light fields emerged as a new imaging modality, enabling to capture all light rays passing through a given amount of the 3D space \cite{Levoy1996}. Compared to traditional 2D images, which only describe the spatial intensity of light rays, captured light fields also describe the angle at which rays arrive at the detection plane. A light field can be represented as a 4D function: $\Omega \times \Pi \to \mathbb{R}, (s, t, u, v) \to L(s, t, u, v)$, where the plane $\Omega$ represents the spatial distribution of rays, indexed by $(u, v)$, while $\Pi$ corresponds to their angular distribution, indexed by $(s, t)$. 3D world coordinates are denoted by $\left( x, y, z \right)$, and for simplicity and without loss of generality, we assume that plane $\Omega$ and $\Pi$ are parallel to the plane of the $x$-$y$ axis.

A common light field operation is to simulate a change of focal length for a traditional photographic camera with a narrow depth of field. 
A ``refocus image'' $I_r$ can be produced through use of the well-known shift-and-sum method~\cite{Ng2005}
\footnote{Note that Equation~\ref{eq:ShiftAndSum} is a re-parameterization of the shift-and-sum equation presented in \cite{Ng2005}, and describes the same operations, despite changes in notation.}, in which the refocus image is obtained as linear combination of shifted light field views:
\begin{equation}
\label{eq:ShiftAndSum}
    I_r(u, v) =
    \sum_{s,t} A\big(s,\, t \big)\, L\big(s,\, t,\, u + (s - s_{r}) \delta ,\, v + (t - t_{r}) \delta \big),
\end{equation}
\noindent where $(s_{r}, t_{r})$ are the indices of the view for which refocus will be performed (``reference view''), $A$ is a filter that defines the synthetic aperture, and $\delta$ is a disparity value, which is related to the refocus distance.

To perform shit-and-sum refocus, it is only necessary to specify $\delta$ as an input parameter. Because the relationship between disparity and refocus distance is not necessarily known, a user may need to use guess-and-check methods to refocus to a specific plane: entering a disparity, viewing the refocus result, and repeating until the desired result is achieved.
However, more intuitive interfaces incorporate disparity information, computed from the light field, to allow the user to specify disparity by clicking on the object to be focused on. \nocite{Sabater2017}

When using the shift-and-sum method described above, the refocus plane is parallel to the light field planes $\Omega$ and $\Pi$ (or ``frontoparallel''). Other refocus methods have been described that allow for non-frontoparallel, planar refocus \cite{Isaksen2000, Vaish2005, Sugimoto2008, Xiao2018}. With these, the results of traditional tilt-shift photography can be simulated, which we refer to as ``tilt-shift refocus.'' Most of these methods utilize planar homographies \cite{Isaksen2000, Vaish2005, Sugimoto2008} to achieve a result and require the guess-and-check for the placement of the refocus plane by specifying input parameters \cite{Isaksen2000, Vaish2005, Xiao2018}. 

A tilted refocus plane has more degrees of freedom than a frontoparallel plane, so it is easier for a user to become confused with the plane's placement, with respect to the scene geometry. The method by Sugimoto and Okutomi allows the user to select a region of interest to focus on \cite{Sugimoto2008}. However, the tilted refocus plane is a side effect applying the homography that optimises the sharpness in the region. As such, the results outside the region can be unpredictable.
Overall, the existing literature does not provide an instructive description of how homographies can be applied to perform tilt-shift refocus in the contemporary light field framework and lack intuitive user specification of the refocus plane.

In this paper we describe a generalization of shift-and-sum that allows for non-frontoparallel refocus planes
, with frontoparallel refocus as a specific case.
This generalized refocus is applied to create a tilt-shift refocus tool that allows for intuitive specification of the refocus plane, visualizing it relative to the a point cloud of the scene geometry. These visualizations are enabled by the inclusion of depth information and camera calibration parameters. With this tool, the user can specify a tilted refocus plane through mouse selection and adjust the parameters with keyboard commands. An example of this can be seen in Figure~\ref{fig:Painter}. Finally, we show that interactive perspective shift can be performed for intermediate views within the generalized shift-and sum framework.

\section{Theory}
\label{sec:Theory}

\begin{figure}
    \centering
    \includegraphics[width=\columnwidth]{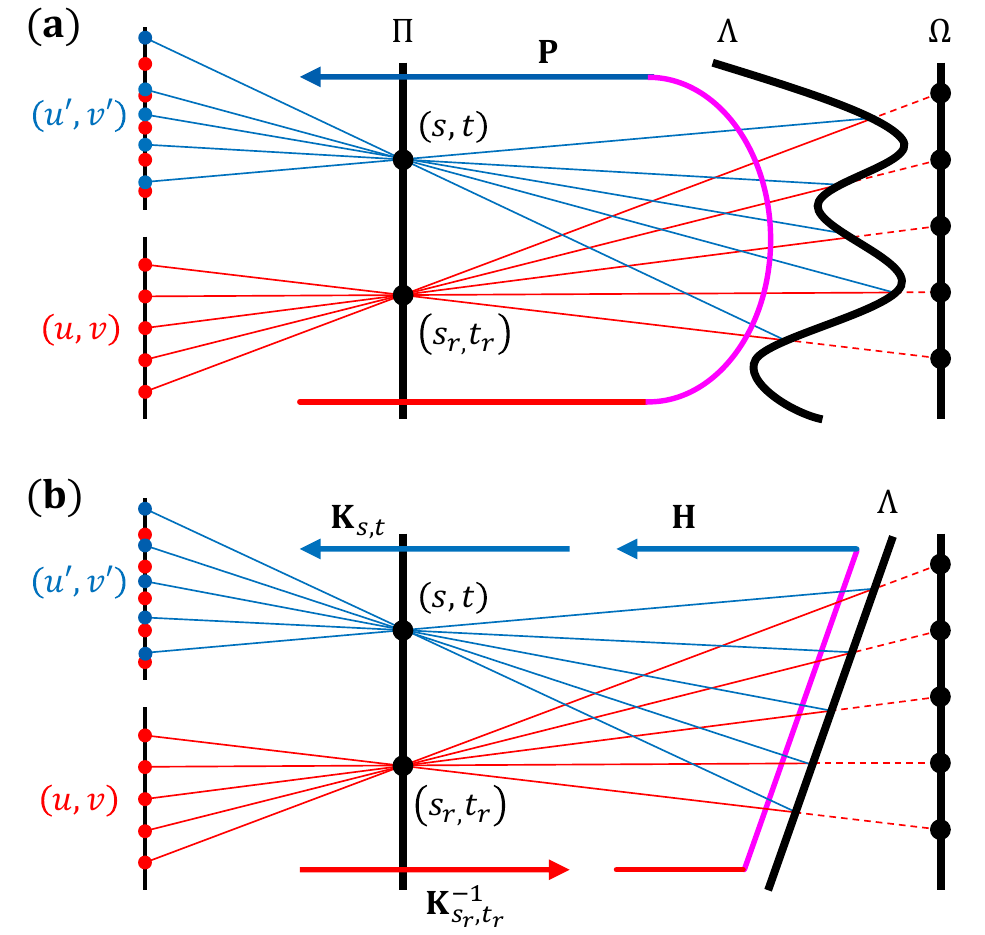}
    \caption{2D visual representation of: \textbf{(a)} transformation of coordinates $\mathbf{P}$ from reference view $\left( s_r, t_r \right)$ to some other view $\left( s, t \right)$, mediated by a refocus surface $\Lambda$; \textbf{(b)} transformation mediated by a planar refocus surface $\Lambda$. Red lines are rays originally sampled at $\left( u, v \right)$, and blue lines are the transformed rays that determine the sampling grid to be interpolated into $\left( u', v' \right)$. Note the irregular sampling $\left( u', v' \right)$ in \textbf{a} and the sampling that has been warped, with respect to the red dots, in \textbf{b}.}
    \label{fig:Transformation}
\end{figure}

\subsection{Generalized shift-and-Sum}

To generalize the basic shift-and-sum method, disparity terms $\delta$ are replaced by a generalized transformation $\mathbf{P}$, specific to the refocus surface $\Lambda$. Then, the refocus image $I_r$ can be expressed as
\begin{equation}
\label{eq:GeneralizedShiftSum}
\begin{split}
    I_r(u,\, v) = \sum_{s,t} A(s,\, t)\, L(s,\, t,\, u',\, v')&,\\
    \text{with} \quad (u',\, v') = \mathbf{P}_{s,t}(u,\, v)&.
\end{split}
\end{equation}
This is the result of reprojection $\mathbf{P}_{s,t}$ of pixel coordinates $\left( u, v \right)$ in the reference view $\left(s_r, t_r\right)$ onto the refocus surface $\Lambda$, followed by projection into view $\left( s, t \right)$. This is represented visually in Figure~\ref{fig:Transformation}a. Note that primed superscripts on variables do not have the same meaning as in \cite{Ng2005}.


\subsection{Tilt-Shift Refocus}

While the application of homographies to produce tilted refocus planes has been mentioned previously \cite{Isaksen2000, Vaish2005, Sugimoto2008}, we provide details below, in the generalized shift-and-sum formalism, for completeness and instructiveness.

For the specific case of planar refocus, we consider a plane $\Lambda$, described by a point $\mathbf{p}$ normal vector $\mathbf{n}$. Then, the projection $\mathbf{P}_{s,t}$ can be described in terms of a planar homography $\mathbf{H}_{s,t}$ and camera intrinsic matrices $\mathbf{K}_{s,t}$ through
\begin{equation}
\label{eq:Projection}
    \mathbf{P}_{s,t} = \mathbf{K}_{s,t} \mathbf{H}_{s,t} \mathbf{K}_{s_r,t_r}^{-1},
\end{equation}
with
\begin{equation}
\label{eq:Homography}
    \mathbf{H}_{s,t} = \mathbf{R}_{s,t} - \frac{\mathbf{t}_{s,t} \mathbf{n}^T}{d}
\end{equation}
and
\begin{equation}
\label{eq:dist_plane_cam}
    d = (\mathbf{p} - \mathbf{t}_{s_r,t_r}) \mathbf{n}^T,
\end{equation}
where $\mathbf{R}_{s,t}$ is the rotation matrix of camera $\left( s, t \right)$; $\mathbf{t}_{s,t}$ is the translation vector between cameras $\left( s, t \right)$ and $\left( s_r, t_r \right)$; and $d$ is the distance from the reference camera $\left( s_r, t_r \right)$ to plane $\Lambda$. A visual representation of this transformation is shown in Figure~\ref{fig:Transformation}b. It is assumed that all camera parameters, intrinsic and extrinsic, are known. The refocus plane normal vector $\mathbf{n}$, and point $\mathbf{p}$ can be specified directly or, in our case, interactively by the user.



In practice, the refocus is not done in a pointwise manner as implied by Equation~\ref{eq:GeneralizedShiftSum}. Instead, the transformation $\mathbf{P}_{s,t}$ can be applied to warp a whole view at once because there is no dependence of $\mathbf{P}_{s,t}$ on $\left(u, v \right)$. Then, similar to shift-and-sum, the weighted average of all views is taken, with a mask applied to each to avoid contributions from empty pixels outside the bounds of the parallelogram containing the warped image.


It can be shown that frontoparallel refocus is a special case of tilt-shift refocus. If we assume that all camera focal lengths are identical, cameras have parallel optical axes ($\mathbf{R}_{s, t} = \mathbf{I})$, and the normal is parallel to the optical axes ($\mathbf{n}^T = \left[ 0, 0, 1 \right]$), Equation~\ref{eq:GeneralizedShiftSum} takes the form of Equation~\ref{eq:ShiftAndSum} when $\mathbf{P}$ is of the form in Equation~\ref{eq:Projection}.


\subsection{Intermediate View Perspective Shift}
\label{sec:PerspectiveShift}

As is the case for the original shift-and-sum algorithm, the angular coordinates $\left(s_r,\, t_r\right)$ of the reference view do not need to coincide with the angular coordinates of a discrete light field view \cite{Ng2005}.
Choosing an intermediate $\left(s_r,\, t_r\right)$ can thus produce a refocus image simulating capture by a camera at a position between the real cameras. 
However, this effect will only be compelling if, when moving the virtual camera position $\left(s_r,\, t_r\right)$, the set of light field cameras included in the aperture $A$ is updated, \textit{i.e} new angular information is taken into account.

\section{Methods}
\label{sec:Methods}

We propose three different methods to intuitively and interactively define the refocus plane parameters $\mathbf{p}$ and $\mathbf{n}$. Two of these methods require a depth map, and all of them include use depth information to provide the user with a visualization of the position of the refocus plane, relative to the scene geometry.

To create a point cloud of a scene, disparity is converted to metric depth as per \cite{Wanner2013}. The pixels $\left(u, v \right)$ in each view $\left(s, t\right)$ are converted from camera coordinates to 2D, homogeneous world coordinates $\left(u_w, v_w \right)$ by reprojection via $\mathbf{K}^{-1}_{s,t}$. Then, these 2D coordinates are converted to 3D world coordinates by multiplying by depth $z$. This is expressed as
\begin{equation}
\label{eq:Reprojection}
    \begin{bmatrix} x \\ y \\ z \end{bmatrix} 
    = z \mathbf{K}^{-1}_{s,t}
    \begin{bmatrix}
    u \\ v \\ 1
    \end{bmatrix}.
\end{equation}

\subsection{Single-Click Definition}
\label{sec:SingleClick}

In the most simple interactive refocus plane definition method, the user defines the refocus plane by selecting, with a mouse, a single point in the reference view (Figure~\ref{fig:OnePoint}, left). Invisible to the user, the normal map of the scene has been created from the point cloud information using \cite{Hoppe1992}. Point $\mathbf{p}$ is calculated, using Equation~\ref{eq:Reprojection}, from the pixel $\left( u, v\right)$ selected, and the corresponding normal vector $\mathbf{n}$. Once these refocus plane parameters have been obtained, $\Lambda$ can be visualized in the point cloud. 

\subsection{Three-Click Definition}
\label{sec:ThreeClick}
A second refocus plane definition method has the user select points to define a plane. The user is presented with a rendering of the point cloud, which they can manipulate as needed before selecting three points (Figure~\ref{fig:ThreePoint}, left). Any of the selected points can serve as $\mathbf{p}$, and the normal $\mathbf{n}$ is found from the cross-product of the vectors between the points. Though this method has a few more steps than the single-click method, it can be used to force the refocus plane through multiple, disparate objects, whereas the single-click method is limited to refocus planes with normals in the normal map.

\subsection{Keyboard Definition}
\label{sec:Keyboard}
Where the methods given above require depth information to function, the final method has the user define the refocus plane manually with keyboard commands, and the point cloud is only used to visualize the placement. The point $\mathbf{p}$ is assumed to lie on the optical axis of the reference camera, with the user specifying the distance $z$. The normal vector $\mathbf{n}$ is not specified directly. Instead, the user specifies the plane's rotation about the Cartesian axes, since this is more intuitive. The visual representation of the plane is updated as the user steps through different values of distance and angle so that it is easy to see how the refocus result will relate to the geometry of the scene, via the point cloud.

While keyboard definition may require more steps to produce the desired refocus plane placement, it is possible to define any plane by using it. In contrast, the click-based methods described in Sections~\ref{sec:SingleClick}~\&~\ref{sec:ThreeClick} are limited by points that can be selected from the scene. For this reason, it is useful to include keyboard definition as a second step that follows either of the mouse-based methods to allow for fine adjustment, in case the original result was not exactly as desired.
\section{Results}
\label{sec:Results}

\begin{figure}
    \begin{subfigure}[]{0.49\columnwidth}
        \centering
        \includegraphics[width=\linewidth]{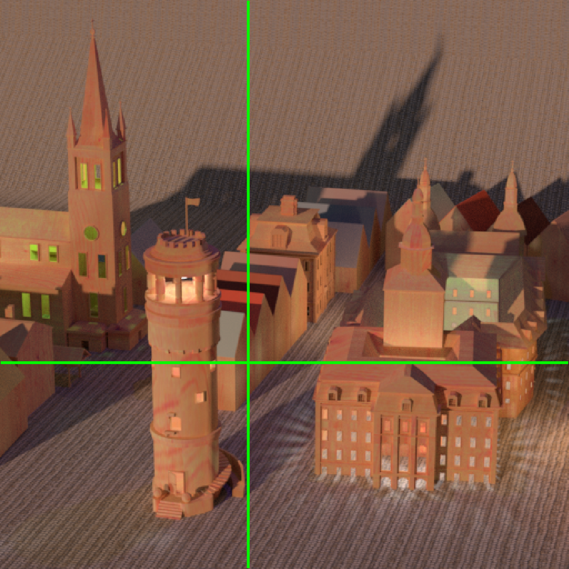}
    \end{subfigure}
    \begin{subfigure}[]{0.49\columnwidth}
        \centering
        \includegraphics[width=\linewidth]{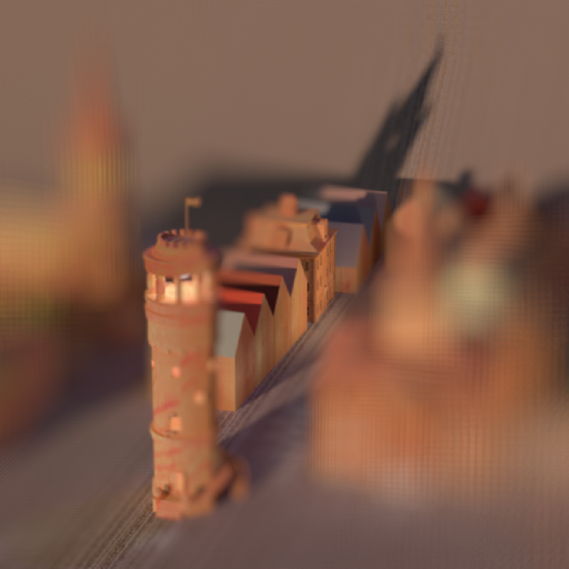}
    \end{subfigure}
    
    \begin{subfigure}[]{0.49\columnwidth}
        \centering
        \includegraphics[width=\linewidth]{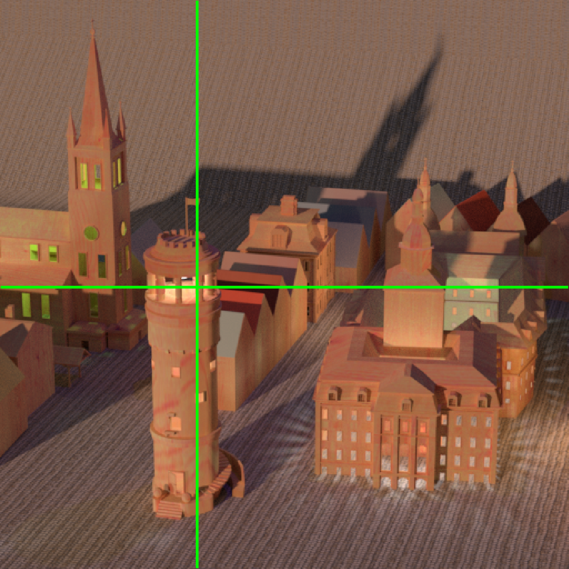}
    \end{subfigure}
    \begin{subfigure}[]{0.49\columnwidth}
        \centering
        \includegraphics[width=\linewidth]{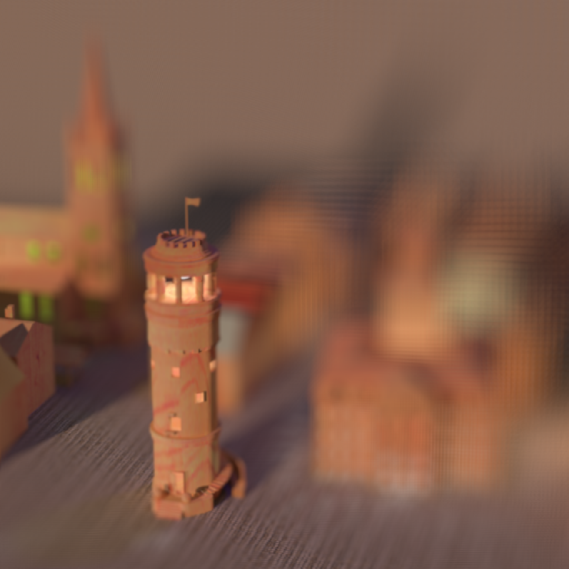}
    \end{subfigure}
    
    \begin{subfigure}[]{0.49\columnwidth}
        \centering
        \includegraphics[width=\linewidth]{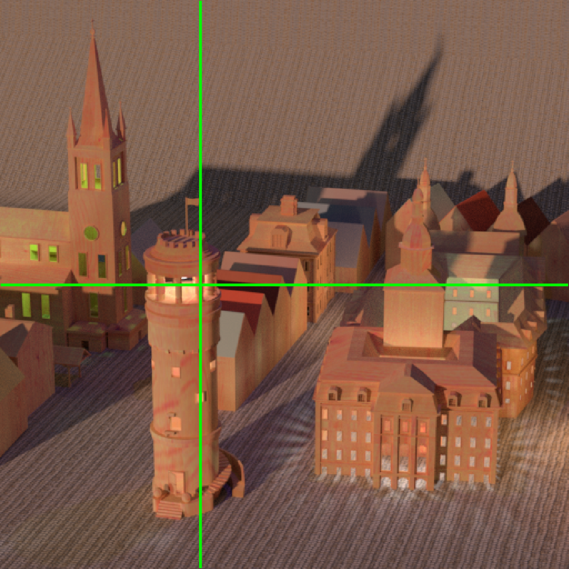}
    \end{subfigure}
    \begin{subfigure}[]{0.49\columnwidth}
        \centering
        \includegraphics[width=\linewidth]{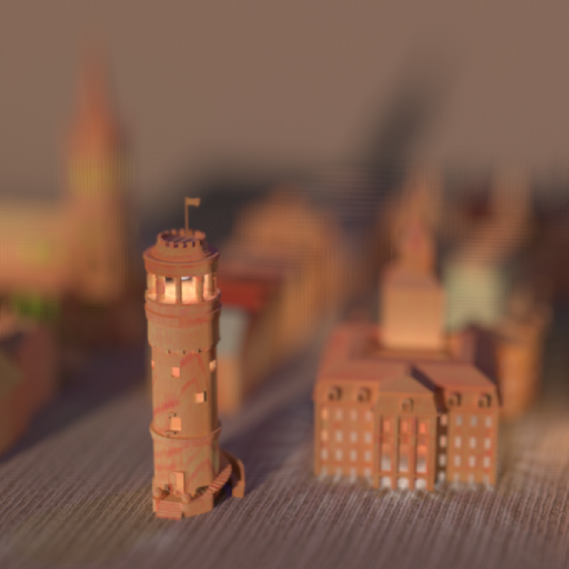}
    \end{subfigure}
    \caption{Tilt-shift refocus images produced with single-click interactive refocus on the ``Tower'' light field from the HCI dataset \cite{honauer2016}. \textbf{(left)} Single-click selection on the reference view, with selected point at intersection of green crosshairs. \textbf{(right)} Tilt-shift refocus results for each selection in the left column. Scene geometry of the refocus plane placement, similar to Figure~\ref{fig:Painter}, top, is available to the user but not shown here.  Image size has been reduced in this pre-print.}
\label{fig:OnePoint}
\end{figure}

\begin{figure}[t!]
    \begin{subfigure}[]{0.495\columnwidth}
        \centering
        \includegraphics[width=\columnwidth, height=2.9cm, keepaspectratio]{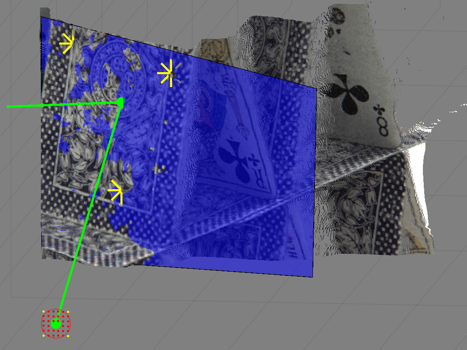} 
    \end{subfigure}
    \begin{subfigure}[]{0.495\columnwidth}
        \centering
        \includegraphics[width=\columnwidth, height=2.9cm, keepaspectratio]{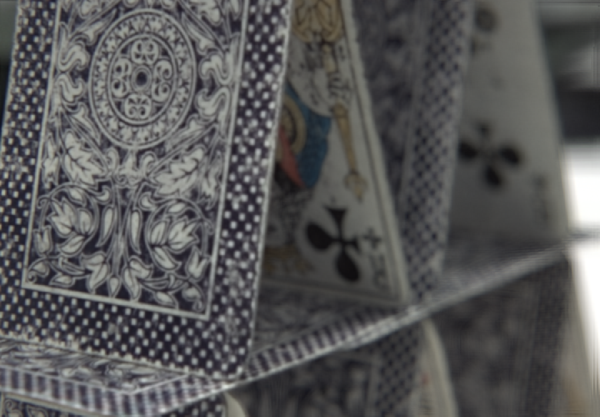}
    \end{subfigure}
    
    \begin{subfigure}[]{0.495\columnwidth}
        \centering
        \includegraphics[width=\columnwidth, height=2.9cm, keepaspectratio]{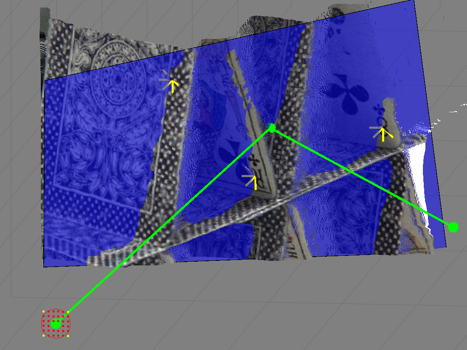}
    \end{subfigure}
    \begin{subfigure}[]{0.495\columnwidth}
        \centering
        \includegraphics[width=\columnwidth, height=2.9cm, keepaspectratio]{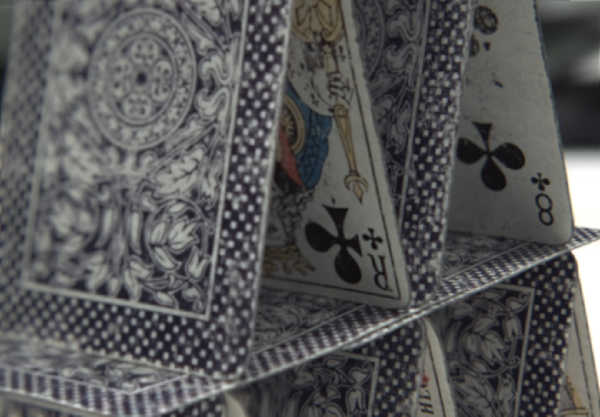}
    \end{subfigure}
\caption{Tilt-shift refocus images produced with three-click interactive refocus on a light field captured with a calibrated Lytro Illum. \textbf{(left)} Point clouds of the scene geometry with selected points as yellow stars ($	\ast$), the refocus plane in transparent blue, camera positions in red and yellow, and optical axis and normal vector in green. \textbf{(right)} Tilt-shift refocus results for each selection in the left column.  Image size has been reduced in this pre-print.}
\label{fig:ThreePoint}
\end{figure}

\begin{figure}[t!]
    \begin{subfigure}[]{0.56\columnwidth}
        \centering
        \includegraphics[width=\linewidth, height=3.5cm, keepaspectratio]{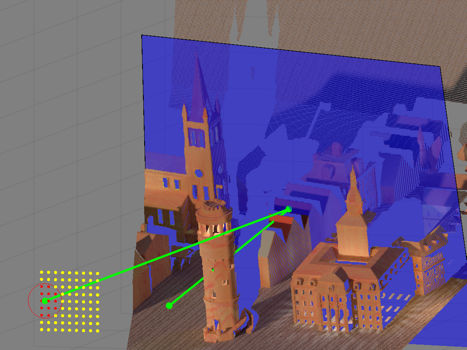} 
    \end{subfigure}
    \begin{subfigure}[]{0.42\columnwidth}
        \centering
        \includegraphics[width=\linewidth, height=3.5cm, keepaspectratio]{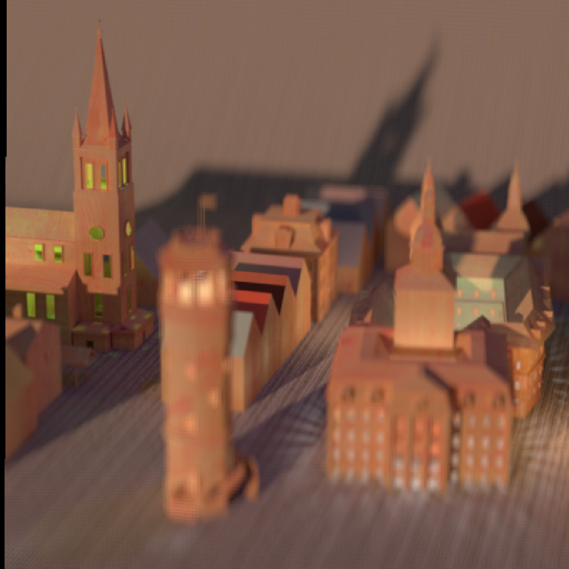}
    \end{subfigure}
    
    \begin{subfigure}[]{0.56\columnwidth}
        \centering
        \includegraphics[width=\linewidth, height=3.5cm, keepaspectratio]{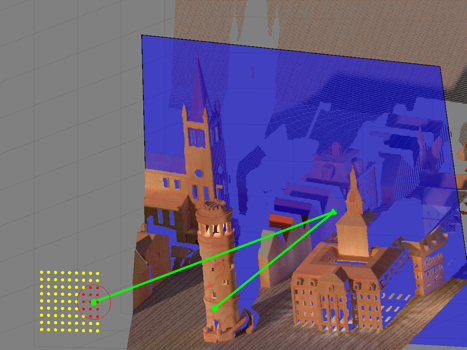}
    \end{subfigure}
    \begin{subfigure}[]{0.42\columnwidth}
        \centering
        \includegraphics[width=\columnwidth, height=3.5cm, keepaspectratio]{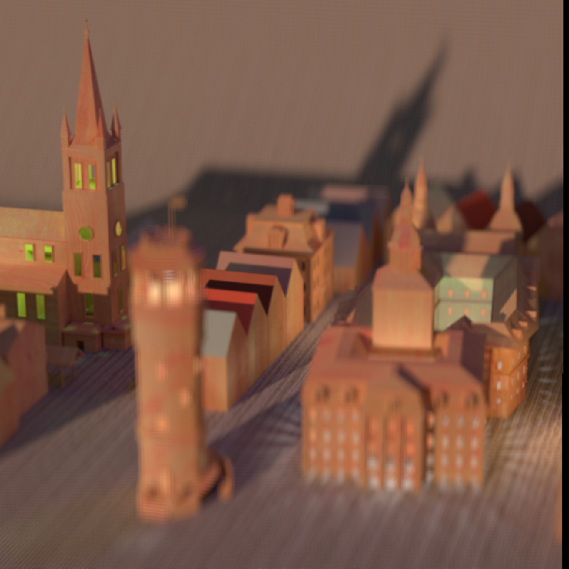}
    \end{subfigure}
\caption{Refocus images for intermediate perspectives, created from the "Tower" light field from the HCI dataset \cite{honauer2016}. \textbf{(left)} Point clouds of the scene geometry, with cameras in the active aperture indicated in red. \textbf{(right)} Perspective shift results for each of the apertures in the left column. Image size has been reduced in this pre-print.}
\label{fig:Shift}
\end{figure}

Example tilt-shift refocus results are shown in Figures \ref{fig:Painter}, \ref{fig:OnePoint}, \ref{fig:ThreePoint}, and \ref{fig:Shift}. Additional results can be found online\footnote{\url{https://v-sense.scss.tcd.ie/research/tilt-shift/}}.

Figure~\ref{fig:OnePoint} shows results of the single-click interactive definition discussed in Section~\ref{sec:SingleClick}. For simulated scenes, such as the one shown, estimated surface normals are quite clean. This means clicking on a planar surface, as shown in the upper-left, will produce a refocus plane across that surface. Results, away from the point clicked, can be more unpredictable for complex surfaces and real scenes, where the normal maps have noise. The middle and bottom rows of Figure~\ref{fig:OnePoint} show how a small difference in the point selected can produce quite different refocus planes. 

Figure~\ref{fig:ThreePoint} shows results of the three-click method, discussed in Section~\ref{sec:ThreeClick}, for a scene captured with a calibrated plenoptic camera. Though requiring more input from the user, the three-click method is more resilient to noise and allows for planes spanning multiple surfaces to be defined more easily, such as in the bottom row of Figure~\ref{fig:ThreePoint}. Where the single-click method is subject to errors in the depth map and normal estimation, the three-click method is only subject to errors in the depth map.

Figures~\ref{fig:Painter}~\&~\ref{fig:Shift} show results of the interactive keyboard definition discussed in Section~\ref{sec:Keyboard}. While it has the most complicated controls and requires more steps to achieve the desired result, compared to the other methods, this method is the most robust. The only dependence of the method on the depth map is in the visualization of the refocus plane, relative to the point cloud. As mentioned previously, it can be beneficial to make a first estimate with single or three-click methods and, then, refine with keyboard definition. Control of the virtual aperture size and position, Section~\ref{sec:PerspectiveShift}, can be provided with the keyboard interface, with an example shown in Figure~\ref{fig:Shift}.

\section{Conclusion}
\label{sec:Conclusion}

Here we have provided formalism for light field tilt-shift refocus in a generalized shift-and-sum framework. We have demonstrated interactive capabilities, enabled by inclusion of depth information, that had not been previously considered and provided a qualitative analysis of the benefits and drawbacks of three different refocus plane definition methods. 

Currently, refocus images from light fields with large separation between cameras contain significant angular aliasing artifacts (ex: Figure~\ref{fig:Painter}). Addressing this problem, either through filtering or view interpolation, is the focus of current work.

While we have discussed only one specific case of generalized shift-and-sum (planar refocus), it should be possible to simulate other refocus surfaces in this framework, similar to \cite{Xiao2018}. We are also investigating refocus surfaces composed of multiple planes, as a hybrid of tilt-shift refocus and generalized refocus surfaces.

\bibliographystyle{IEEEbib.bst}
\bibliography{ELFI_full.bib}

\end{document}